\documentclass[reprint,prb,longbibliography]{revtex4-1}
 \usepackage{amsmath,amssymb}
 \usepackage{stmaryrd}
 \usepackage{latexsym}
 \usepackage{CJK}
 \usepackage{graphicx}
 \usepackage{indentfirst}
 \usepackage{listings}
 \usepackage{xcolor}
 \usepackage{booktabs}
 \usepackage{stmaryrd}
 \usepackage{multirow}
 \usepackage{verbatim}
 \usepackage{makecell}
 \usepackage{lineno}
 \usepackage{longtable}
 \usepackage{hyperref}
 
 \hypersetup{colorlinks,breaklinks,
            linkcolor=blue,urlcolor=blue,anchorcolor=blue,citecolor=blue}

 \newcommand{\uc}{\text{c}}       
 \newcommand{\ud}{\text{d}}
 \newcommand{\ue}{\text{e}}
 \newcommand{\ug}{\text{g}}
                      
 \newcommand{\un}{\text{n}}       
 \newcommand{\us}{\text{s}}       
 \newcommand{\uB}{\text{B}}
 
 \newcommand{\uE}{\text{E}}
 \newcommand{\uG}{\text{G}}
 \newcommand{\uI}{\text{I}}
 \newcommand{\uS}{\text{S}}
 \newcommand{\csch}{\text{csch}}

\begin{document}

\title{On the Temperature Dependence of Grain Boundary Mobility}

\author{Kongtao Chen}
\affiliation{Department of Materials Science and Engineering, University of Pennsylvania, Philadelphia, PA 19104 USA}
\author{Jian Han}
\affiliation{Department of Materials Science and Engineering, University of Pennsylvania, Philadelphia, PA 19104 USA}
\affiliation{Department of Materials Science and Engineering, City University of Hong Kong, Kowloon, Hong Kong, China}
\author{David J. Srolovitz}
\affiliation{Department of Materials Science and Engineering, University of Pennsylvania, Philadelphia, PA 19104 USA}
\affiliation{Department of Mechanical Engineering and Applied Mechanics, University of Pennsylvania, Philadelphia, PA 19104 USA}
\affiliation{Department of Materials Science and Engineering, City University of Hong Kong, Kowloon, Hong Kong, China}  
\date{\today}
\date{\today}

\begin{abstract}
The grain boundary (GB) mobility relates GB velocity to the thermodynamic driving forces and is central to our understanding microstructure evolution in polycrystals. 
Recent molecular dynamics (MD)  and experimental studies have shown that the temperature-dependence of the GB mobility is much more varied than is commonly thought.
GB mobility may increase, decrease, remain constant or show multiple peaks with increasing temperature.
We propose a mechanistic model for GB migration, based on the formation and migration of line defects (disconnection) within the GB.
We implement this model in a kinetic Monte Carlo and statistical mechanics framework; the results capture all of these observed temperature dependences and are shown to be in quantitative agreement with each other and direct MD simulations of GB migration for a set of specific grain boundaries. 
Examination of the dependence of GB mobility on disconnection mode and temperature provides new insight into how GBs migrate in polycrystalline materials.

\begin{description}
\item[Keywords]
Grain Boundary; Mobility; Molecular Dynamics; Monte-Carlo; Thermodynamics
\end{description}

\end{abstract}

\maketitle

\section{Introduction}

Our current understanding of microstructural evolution in polycrystalline materials is based, in large part, on our understanding of how grain boundaries (GBs) move. 
Normally, we describe GB dynamics as overdamped, where the GB velocity $v$ is proportional to the driving force $F$ (the variation of free energy with respect to GB displacement), $v=MF$, where the proportionality constant $M$ is the GB mobility~\cite{Turnbull1951}.
Since both the force and velocity are vectors, the mobility is a second rank tensor~\cite{Chen2020}. 
Here, we focus on the normal components of the GB velocity and  driving force and the component of the mobility tensor that couples them (i.e., here, $M$ is a scalar).
Since the most widely employed approach for controlling  the rate (and often the nature) of microstructure evolution is through variation of temperature $T$ (i.e., annealing), this study focuses on the temperature dependence of the GB mobility. 

The temperature-dependence of grain boundary mobility has been measured for a wide range of materials both  experimentally~\cite{Aust1959a,Aust1959, Rutter1965, Gottstein2009} and via atomistic simulations~\cite{Homer2014, Rahman2014a,Janssens2006,Zhang2004,Priedeman2017,Zhang2005,Olmsted2009b,Zhou2011,Song2012,Upmanyu1999,Schonfelder2005}.
The quoted references focused on the measurement of the mobility of nominally flat GBs in bicrystals of elemental metals rather than GBs in microstructures (i.e., averaging over many GBs or influenced by GB junctions).
The temperature-dependence of the GB mobility is commonly fit to an Arrhenius relation $M=M_0 e^{-Q/k_\uB T}$, where $Q$ is an activation energy, the prefactor $M_0$ is a  constant, and $k_\uB$ is the Boltzmann constant.
This Arrhenius relation  provides a good fit to many of the $M$ vs. $T$  experimental data. 
However, GB mobilities extracted from atomistic simulations  in pure systems shows a wide variety of 
$T$-dependences~\cite{Homer2014,Olmsted2009b,Schonfelder2005}. 
Homer et al.~\cite{Homer2014} performed a series of MD simulations of GB migration in Ni for a  large number of bicryallographically different GBs.
While nearly half of these  GBs showed mobilities that they characterized as Arrhenius over some range of temperature, they also observed several/many cases for which the GB mobility  (i) decreased with increasing $T$ (so called anti-thermal  behavior), (ii) was nearly $T$ independent,  (iii) exhibit maxima and/or minima with respect to $T$, and (iv) is nearly zero at low $T$ and then increases rapidly over a small $T$-range.
The existence of such diversity in GB mobility $M(T)$ within a single material challenges our current understanding of and  ability to predict GB mobility.

Recent studies suggest that GB migration occurs through the motion of line defects (i.e., disconnections), that are constrained to lie within the GB and are characterized by both a Burgers vector $\mathbf{b}$ and a step height $h$ that are determined  by the underlying GB bicrystallography~\cite{Hirth1973,Pond1979,King1980,Balluffi1982,Hirth2006,Hirth2007,Cahn2006b,Thomas2017,Han2018,Chen2019}. 
Both atomistic simulation~\cite{Rajabzadeh2013a, Khater2012a, Combe2016} and electron microscopy~\cite{Legros2008, Mompiou2009, Rajabzadeh2013} studies have  directly observed GB migration through the formation and migration of disconnections along  GBs. 
The motion of disconnections of different modes, characterized by different $(\mathbf{b}_m,h_m)$ ($m$ is the index for mode) allowed by the bicrystallography, can conspire to affect GB motion. 
This suggests a possible source for some of the complexity in the observed temperature dependence of GB mobility~\cite{Thomas2017,Han2018,Chen2019}.

In this paper, we explore the $T$-dependence of GB mobility based upon kinetic Monte Carlo (kMC) and molecular dynamics (MD) simulations of GB migration. 
We then propose an analytical model to describe many of the observed forms of the temperature dependences of  GB mobility.

\section{Kinetic Monte Carlo Simulations}\label{sec:kmc}
\subsection*{Model}

\begin{figure}
\includegraphics[height=0.8\linewidth]{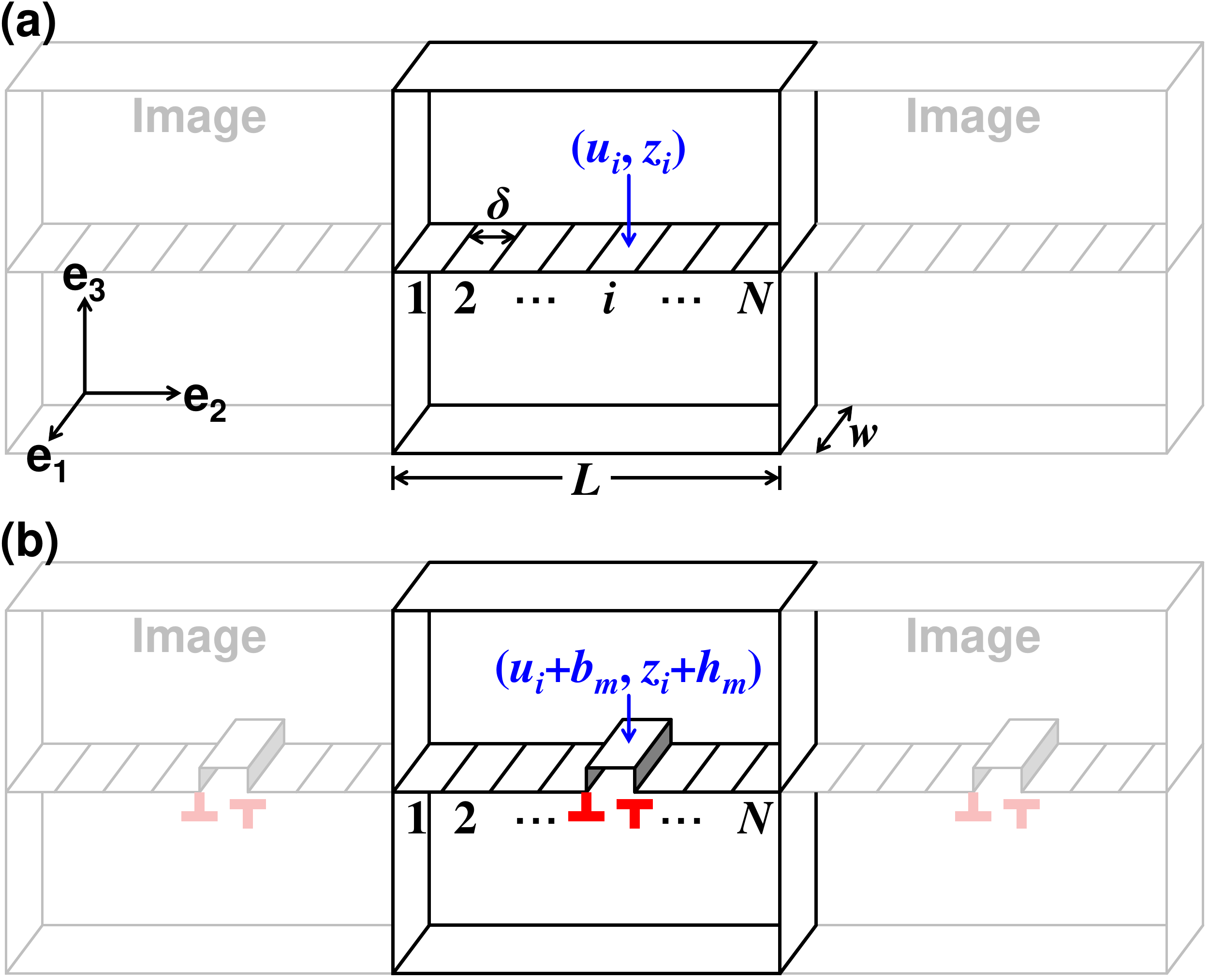}
\caption{\label{KMCmodel}(a) Quasi-1D lattice model description of a tilt GB. The state of  lattice site  $i$ is $(u_i, z_i)$. 
(b) The GB state after nucleation of a pair of disconnections of mode $m$ (i.e., $(b_m, h_m)$ and $(-b_m, -h_m)$) at site $i$).
}
\end{figure}

Here, we describe the motion of a GB in terms of the formation and migration of disconnections, as illustrated in Fig.~\ref{KMCmodel}. 
This model describes a quasi-2D bicrystal, containing a quasi-1D tilt GB. 
The tilt axis of the GB is along the $\mathbf{e}_1$-axis and the GB normal  is $\mathbf{e}_3$ (see Fig.~\ref{KMCmodel}a). 
By ``quasi'', we imply that the  bicrystal structure is uniform along the $\mathbf{e}_1$-axis (i.e., the model is thin in the $\mathbf{e}_1$-direction). 
Periodic boundary conditions are applied in both the $\mathbf{e}_1$- and the $\mathbf{e}_2$-axes. 
The GB is discretized into $N$ lattice sites along the $\mathbf{e}_2$-axis. 
The state of the $i^\text{th}$ lattice site ($1\le i \le N$) on the GB is denoted by $(u_i, z_i)$, where $u_i$ is the relative (tangential) displacement of the upper grain with respect to the lower one (i.e., eigen-shear) in the $\mathbf{e}_2$ direction and $z_i$ is the position of the GB plane in the $\mathbf{e}_3$-direction at GB-site $i$. 
Formation of a $\pm$ disconnection pair of mode $m$, $(b_m, h_m)$ and $(-b_m, -h_m)$, at site $i$ corresponds to the following change: $(u_i, z_i) \rightarrow (u_i+b_m, z_i+h_m)$, as illustrated in Fig.~\ref{KMCmodel}b. 
The evolution of the state of the GB   is described by $\{(u_i(t), z_i(t))\}$ for all GB sites $i$. 
Disconnection motion is represented as transitions in the GB states on a series of sites. 
For example, referring to Fig.~\ref{KMCmodel}b, if the  right disconnection ``$\top$'' glides to the right by one lattice site (a distance $\delta$), the state of site $(i+1)$ changes from $(u_{i+1}, z_{i+1})$ to $(u_{i+1}+b_m, z_{i+1}+h_m)$. 
  
Under a  driving force, the GB lattice site states shown in Fig.~\ref{KMCmodel} will evolve. 
In the kMC simulations described here, we focus on a driving force associated with a difference (or jump) in the chemical potential across the GB, denoted by $\psi$. 
$\psi$ can represent driving forces of  different physical origins.  
For example, $\psi$ can represent a capillary force (curvature), the difference in strain energy between two differently oriented grains in  a bicrystal with anisotropic elastic constants subject to a non-shear stress~\cite{Washburn1952} or the effects of an applied magnetic field in a material with orientation-dependent magnetic susceptibility~\cite{Gunster2013}. 
In atomistic simulations, $\psi$ is often modeled using the synthetic driving force method~\cite{Janssens2006}. 
GB migration can be driven by other types of driving forces, such as a shear stress parallel to the GB plane $\tau$. 
However, in most experiments~\cite{Hu1972,Grunwald1970,Molodov1995,Viswanathan1973,Lejcek1994,Furtkamp1998, Molodov1998} and atomistic simulations~\cite{Olmsted2009b,Homer2014}, GB mobility is measured based upon chemical potential jump driving forces $\psi$ (e.g., curvature-driven GB motion)~\cite{Gottstein2009}.

\subsection*{Algorithm}

In this section, we describe the kinetic Monte Carlo (kMC) algorithm we employ to simulate GB migration under a chemical potential jump driving force $\psi$.

The first step is to determine the Burgers vector and step height for each disconnection mode $m$ $(b_m, h_m)$, the energy landscape and the work done by the external driving force $W^\uE_m \equiv -\psi h_m \delta$, where $\delta$ is the periodicity of the local energy landscape along the GB which we associate with width of the GB site.  We initialize the model by assuming that the GB is disconnection-free along $u_i = 0$, $z_i = 0$ (for $i = 1, \cdots, N$) at time $t = 0$. 
Then, the kMC simulation proceeds as follows.

\begin{figure}
\includegraphics[height=0.54\linewidth]{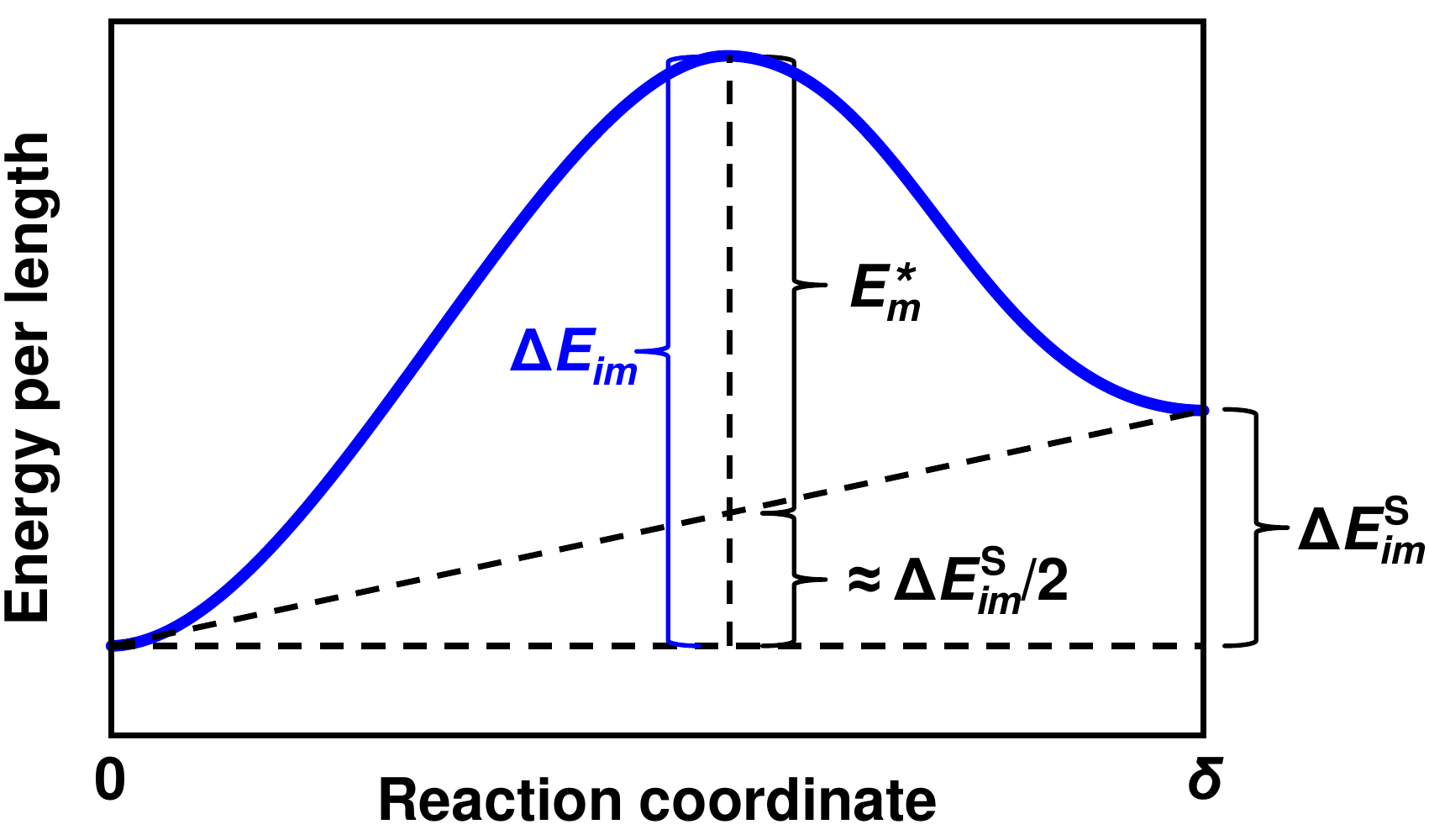}\hspace{-1.78em}%
\caption{\label{KMCbarrier}Schematic of the energy barrier associated with a single transition associated with the formation  of a pair of disconnections of mode $m$ at  site $i$.  }
\end{figure}

\begin{itemize}
\item[(i)] List the energy barriers for all possible transition events. 
The energy change associated with the formation of a disconnection pair of mode $m$ at site $i$  $\Delta E^\uS$ is  (see Fig.~\ref{KMCbarrier}):
\begin{align}\label{DEim}
\Delta E_{im}
&= \frac{1}{2}\Delta E^\uS_{im} + E^\ast_m
\nonumber\\
&= \frac{1}{2} \left(
\Delta E^\uc_{im} + W^\uI_{im} + W^\uE_m
\right)
+ E^\ast_m, 
\end{align} 
where $\Delta E^\uS_{im}$ is the total energy change associated with forming the disconnection dipole and $E^*_m$ is the disconnection glide barrier. $\Delta E^\uc$ is the formation energy of two disconnection cores: 
\begin{align}
\Delta E^\uc_{im}
&= \gamma \Big[
\left|z_i^+ - z_{i-1}\right|
+ \left|z_{i+1} - z_i^+\right| 
\nonumber\\
& - \left|z_i - z_{i-1}\right|
- \left|z_{i+1} - z_i\right|
\Big]
\nonumber\\
&+ \zeta K \Big[
\left(u_i^+ - u_{i-1}\right)^2
+ \left(u_{i+1} - u_i^+\right)^2 
\nonumber\\
& - \left(u_i - u_{i-1}\right)^2
- \left(u_{i+1} - u_i\right)^2
\Big], 
\end{align}
where the first term is associated with the increased GB area (steps) and the second is an estimate of the core energy associated with the disconnection Burgers vectors\cite{Thomas2017,Han2018,Chen2019,Chen2020} $\delta$.  Here, $\gamma$ is the GB energy associated with the step face,  $K = \mu/4\pi(1-\nu)$ ($\mu$ is the shear modulus and $\nu$ is the Poisson's ratio), and $\zeta$ is a constant. 
$W^\uI$ is the contribution to the work done in forming a disconnection pair associated with the internal stress field $\tau_i$ at site $i$ from all other disconnections in the systems (this is the Peach-Koehler (PK) force): 
\begin{equation}\label{WI_im}
W^\uI_{im}
= - \tau_i b_m \delta
+ \frac{2\pi K b_m^2}{N} \cot\left(\frac{\pi}{2N}\right). 
\end{equation}
The second term in this expression is associated with the expansion of the disconnection dipole from size $0$ to $\delta=L/N$ (i.e.,  the elastic interaction between the two disconnections in the dipole.
Equation~\eqref{DEim} can best be understood by reference to Fig.~\ref{KMCbarrier}.

\item[(ii)] Compile a list of the rates of all possible transitions. 
The transition rate associated with the formation of a mode $m$ disconnection dipole at site $i$  is 
\begin{equation}
\lambda_{im}
= \omega \exp\left(-\frac{\Delta E_{im} w}{k_\uB T}\right),  
\end{equation}
where $\omega$ is an attempt frequency and $w$ is the thickness of the bicrystal in direction $\mathbf{e}_1$. 
Note, disconnection migration is simply the formation of a disconnection dipole adjacent to an existing disconnection.
The  ``activity'' of the system is the rate at which any transition occurs, $\Lambda = \sum_m \sum_i \lambda_{im}$. 

\item[(iii)] Choose an event ($im$) to occur in accordance with the probability distribution $p_{im} = \lambda_{im}/\Lambda$.

\item[(iv)] Advance the clock $t:=t + \Delta t$, where $\Delta t = \Lambda^{-1} \ln(\eta^{-1})$ and $\eta \in (0,1]$ is a random number.

\item[(v)] Update the state of site $i'$ as  $u_{i'} := u_{i'} + b_{m'}$, $z_{i'} := z_{i'} + h_{m'}$ and  the stress at each site in the system ($i = 1,\cdots, N$) as
\begin{align}\label{tau_i_update}
\tau_i := \tau_i
+ &\frac{2\pi K b_{m'}}{N \delta} 
\left\{
\cot\left[\frac{\pi}{N}\left(i' - i - \frac{1}{2}\right)\right] \right. \nonumber\\
&- \left. \cot\left[\frac{\pi}{N}\left(i' - i + \frac{1}{2}\right)\right] \right\}.
\end{align} 

\item[(vi)]Return to Step (i). 
\end{itemize}
(Derivations of Eqs.~\eqref{WI_im} and \eqref{tau_i_update} are provided in  Supplemental Material Section I.)

To perform a simulation, we choose a GB (i.e., including specifying the bicrystallography-allowed $(b_m, h_m)$ and their associated energies), a temperature and a driving force $\psi$.
Each kMC simulation is  run for $10^6$ (variable time) steps and measure a statistically-averaged, steady-state GB migration velocity $v$. 
The GB mobility at this temperature is determined from $M = v/\psi$ for $\psi$ sufficiently small that $v \propto \psi$. 
Such simulations are repeated over a range of temperatures to determine  $M(T)$ for the chosen  GB. 

Two sets of kMC simulations were performed using different disconnection parameter sets.
\begin{itemize}
\item[(i)] The first set of simulations were performed to investigate the contributions of different types of disconnections on $M(T)$.
For these simulations, we employ reduced (dimensionless) quantities, labeled by ``$\sim$'': 
$\tilde{h} = h/\delta$, $\tilde{b} = b/\delta$, 
$\tilde{\gamma} = \gamma/(2\pi K\delta)$, 
$\tilde{\psi} = \psi/(2\pi K)$, 
$\tilde{\tau} = \tau/(2\pi K)$, 
$\tilde{T} = k_\uB T/(2\pi K\delta^2 w)$, 
$\tilde{E} = E/(2\pi K\delta^2)$, 
$\tilde{t} = t\omega$, 
and $\tilde{M} = 2\pi KM/(\delta\omega)$. 
This representation  reduces the number of parameters in the simulations (e.g., by scaling out $\delta$, $\omega$ and $K$). 
For these simulations, we choose 
$\tilde{\gamma} = 1$, 
$\tilde{\psi} = 1$, 
$\zeta = 2\pi$, 
and $\tilde{E}^\ast_m = 0.1\tilde{\gamma}(|\tilde{b}_m| + |\tilde{h}_m|)$. 

\item[(ii)] The second set of kMC simulations are performed using parameters determined from atomistic calculations for $\Sigma 17$ $[100]$ $(035)$ and $\Sigma 25$ $[100]$ $(034)$ symmetric tilt GBs in   Al (using the embedded-atom-method potential from~[\onlinecite{Mishin1999}]), as described  in Section~\ref{MDcomparison}.
The admissible  disconnection modes $\{(b_m, h_m)\}$ for these boundaries are as described in~[\onlinecite{Han2018},\onlinecite{Chen2019}].
\end{itemize}
In all of the simulations reported here, we employed $N = 100$ and the site width $\delta$  equal to the coincidence-site-lattice (CSL) cell in the direction parallel to the GB. 

\begin{figure*}[ht]
\includegraphics[height=0.65\linewidth]{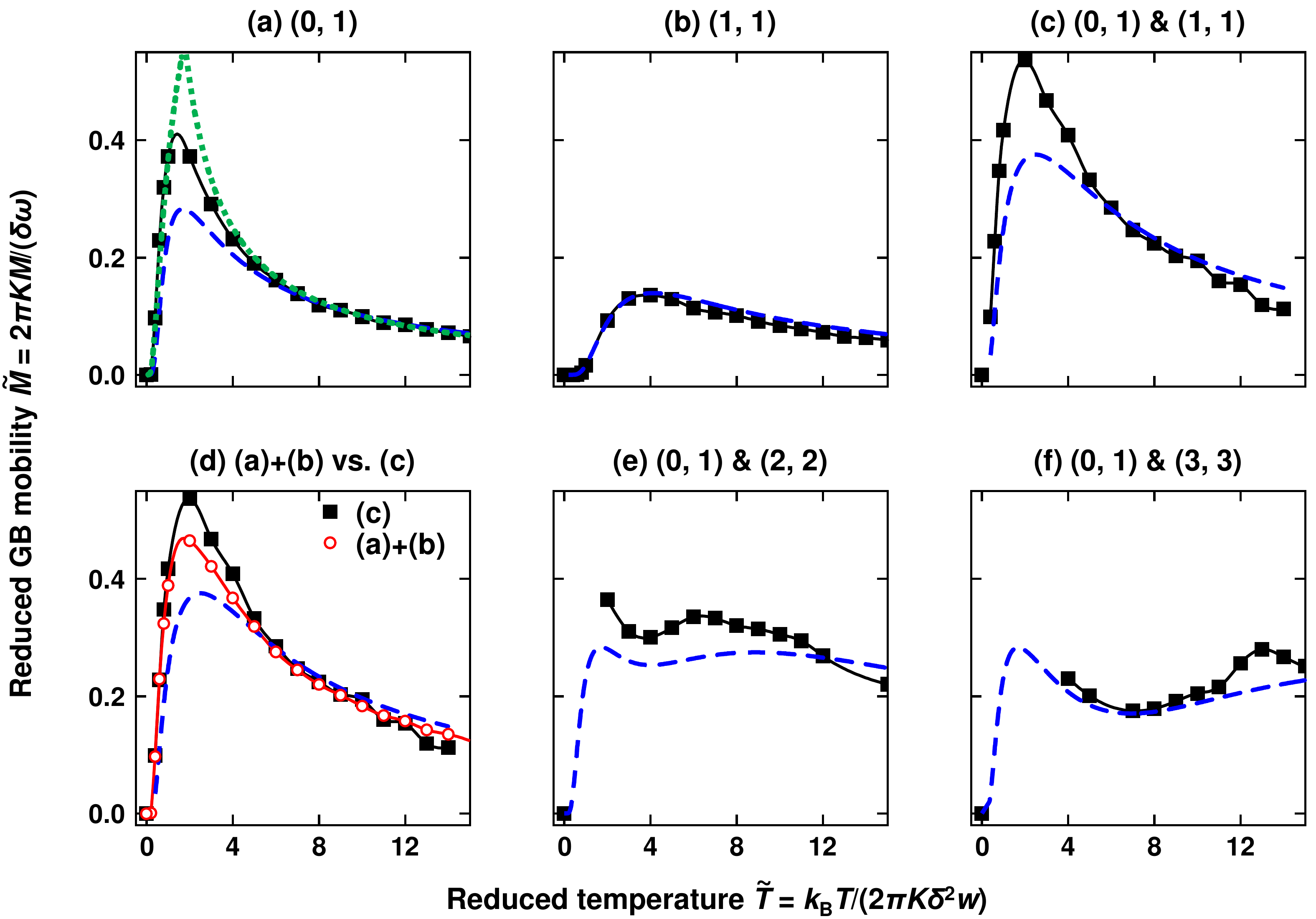}\hspace{-1.78em}%
\caption{\label{fig2}kMC simulation results for the (reduced) temperature dependence of the (reduced) GB mobility. 
The label above each figure indicates the operative disconnection modes: $(\tilde{b}_1, \tilde{h}_1)$ and, if there is another mode, $(\tilde{b}_2, \tilde{h}_2)$ (recall that $\tilde{h}=h/\delta$ and $\tilde{b}=b/\delta$). 
The kMC simulation correspond to: 
(a) a single pure-step disconnection mode -- $(\tilde{b}_1, \tilde{h}_1) = (0,1)$; 
(b) a single mode -- $(\tilde{b}_1, \tilde{h}_1) = (1,1)$; 
(c) two modes -- $(\tilde{b}_1, \tilde{h}_1) = (0,1)$ and $(\tilde{b}_2, \tilde{h}_2) = (1,1)$;
(e) two modes -- $(\tilde{b}_1, \tilde{h}_1) = (0,1)$ and $(\tilde{b}_2, \tilde{h}_2) = (2,2)$; and 
(f) two modes -- $(\tilde{b}_1, \tilde{h}_1) = (0,1)$ and $(\tilde{b}_2, \tilde{h}_2) = (3,3)$. 
(d) shows a comparison between the results from (c) (solid black squares) and the sum of the reduced mobilities for the two 1-mode simulations corresponding to the same two modes, individually, as shown in  (a) and (b) (hollow red circles). This comparison demonstrates that when multiple modes are present, the corresponding mobilities do not simply add.   
In each figure, the kMC results are represented by solid black squares; the solid black lines connecting the kMC data points are drawn as guides to the eye. 
The blue dashed lines are obtained from the analytical model, i.e., Eq.~\eqref{velocity1Model}. 
The green dotted lines in (a) indicate the relationships $\ln \tilde{M} = -1/\tilde{T}$ and $\tilde{M} = 1/\tilde{T}$. 
}
\end{figure*}

\subsection*{Results}\label{sec:kMCresults}

Figure~\ref{fig2} shows the temperature dependence of the reduced GB mobility $\tilde{M}(\tilde{T})$ obtained from the first set of kMC simulations. 
Several cases are investigated in order to understand the effects of the type of the operative disconnection(s) $(b_m, h_m)$ and a  single versus multiple operative disconnections. 

In the first set of kMC simulations, we focus on a single operative disconnection type  that corresponds to a pure-step  $(\tilde{b}_1, \tilde{h}_1) = (0,1)$. 
The results are shown in Fig.~\ref{fig2}a. 
Starting from low $T$, the GB mobility increases quickly with temperature and then decays slowly at hight $T$. 
At low $T$, GB mobility is well described by an  Arrhenius relationship $\ln \tilde{M} \sim -1/\tilde{T}$, while at high $T$  it decays as  $\tilde{M} \sim 1/\tilde{T}$; both of these relations are indicated by the two  dotted green curves in Fig.~\ref{fig2}a). 

In the second kMC simulation, we focus on a single operative disconnection that has both finite  Burgers vector and step height, $(\tilde{b}_1, \tilde{h}_1) = (1,1)$. 
These kMC results are shown in Fig.~\ref{fig2}b. 
The presence of a finite Burgers vector lowers the GB mobility relative to the pure step case. 
Like in the first kMC simulation, the GB mobility $\tilde{M}(\tilde{T})$ increases at low $T$ and then decays at high $T$. 
Two major differences are that the maximum mobility has decreased by more than a factor of two  and that instead of rising rapidly from $\tilde{T}=0$, the GB mobility remains nearly zero until a critical temperature before its initial rise (\textit{cf.}
(the  $(\tilde{b}_1, \tilde{h}_1) = (0,1)$ case in Fig.~\ref{fig2}a).

Three additional kMC simulations shown are performed, corresponding to  two operative disconnection modes.  
In all cases, the first mode is the pure step disconnection $(\tilde{b}_1, \tilde{h}_1) = (0,1)$, while the second has both finite step height and Burgers vector; i.e., $(\tilde{b}_2, \tilde{h}_2) = (1,1)$, $(2,2)$, and $(3,3)$ (see Figs.~\ref{fig2}c,e,f, respectively).
Comparison of these three, 2-mode cases show a much wider range of behavior than in the single mode cases.

While the 2-mode $(\tilde{b}_1, \tilde{h}_1) = (0,1)$,  $(\tilde{b}_2, \tilde{h}_2) = (1,1)$ case (Fig.~\ref{fig2}c) resembles that for the pure step mode case in Fig.~\ref{fig2}a, the highest mobility in the 2-mode case exceeds that in the pure step case by nearly 50\% and that of the finite Burgers vector and step height case of Fig.~\ref{fig2}b by nearly 300\%.
Figure~\ref{fig2}d compares this 2-mode case ($(\tilde{b}_1, \tilde{h}_1) = (0,1)$,  $(\tilde{b}_2, \tilde{h}_2) = (1,1)$) with the superposition of the mobilities of the two 1-modes cases ($(\tilde{b}_1, \tilde{h}_1) = (0,1)$ in Fig.~\ref{fig2}a and  $(\tilde{b}_2, \tilde{h}_2) = (1,1)$ in Fig.~\ref{fig2}b).
Clearly, the mobility in the 2-mode case in not simply the sum of those for the two 1-mode cases.

The 2-mode cases with $(\tilde{b}_2, \tilde{h}_2) = (2,2)$ and $(\tilde{b}_2, \tilde{h}_2) = (3,3)$ both decay at large temperature, like in the other cases, but also exhibit clear minima at intermediate temperatures and maxima at both high and low temperature. 
The minimum is deeper for the $(\tilde{b}_2, \tilde{h}_2) = (3,3)$ mode than for the $(\tilde{b}_2, \tilde{h}_2) = (2,2)$ mode.

\section{Statistical disconnection model} 
\subsection*{Model}

In this section, we develop an expression for the GB mobility 
\begin{equation}\label{defM}
M = \frac{\ud v}{\ud \psi}\bigg|_{\psi=0}
\end{equation}
via a statistical analysis of the formation and migration (glide) of disconnections of one or more modes and compare with the kMC results.  
The first step is predicting the GB migration velocity $v$ as a function of $\psi$. 
For consistency with the kMC simulations, we focus on the quasi-1D GB model  of Fig.~\ref{KMCmodel}. 

We begin by considering GB migration in terms of the formation and glide of a single mode of disconnection $(b,h)$ (we omit the subscript ``$m$'' for a one mode case) on  an infinitely long, initially flat GB under the influence of a driving force $\psi$.  
Statistically, disconnection pair centers will   form uniformly distributed along the GB, with an average spacing between disconnection pairs $L$ (to be determined). 
For simplicity, focus on a periodic distribution of such  pairs (as in  Fig.~\ref{KMCmodel}b);  unlike in the kMC model, here  we  consider a continuous (rather than discrete) GB  and $L$ is a temperature-dependent correlation (rather than fixed) length. 
Under the influence of driving force $\psi$, the two disconnections in each pair glide apart. 
Once the separation between disconnection pairs reaches $L$, they annihilate with disconnections from neighboring pairs; at this point, the GB  becomes flat again, but has migrated a distance $h$. 

\subsection*{Energetics}

\begin{figure}
\includegraphics[height=1.4\linewidth]{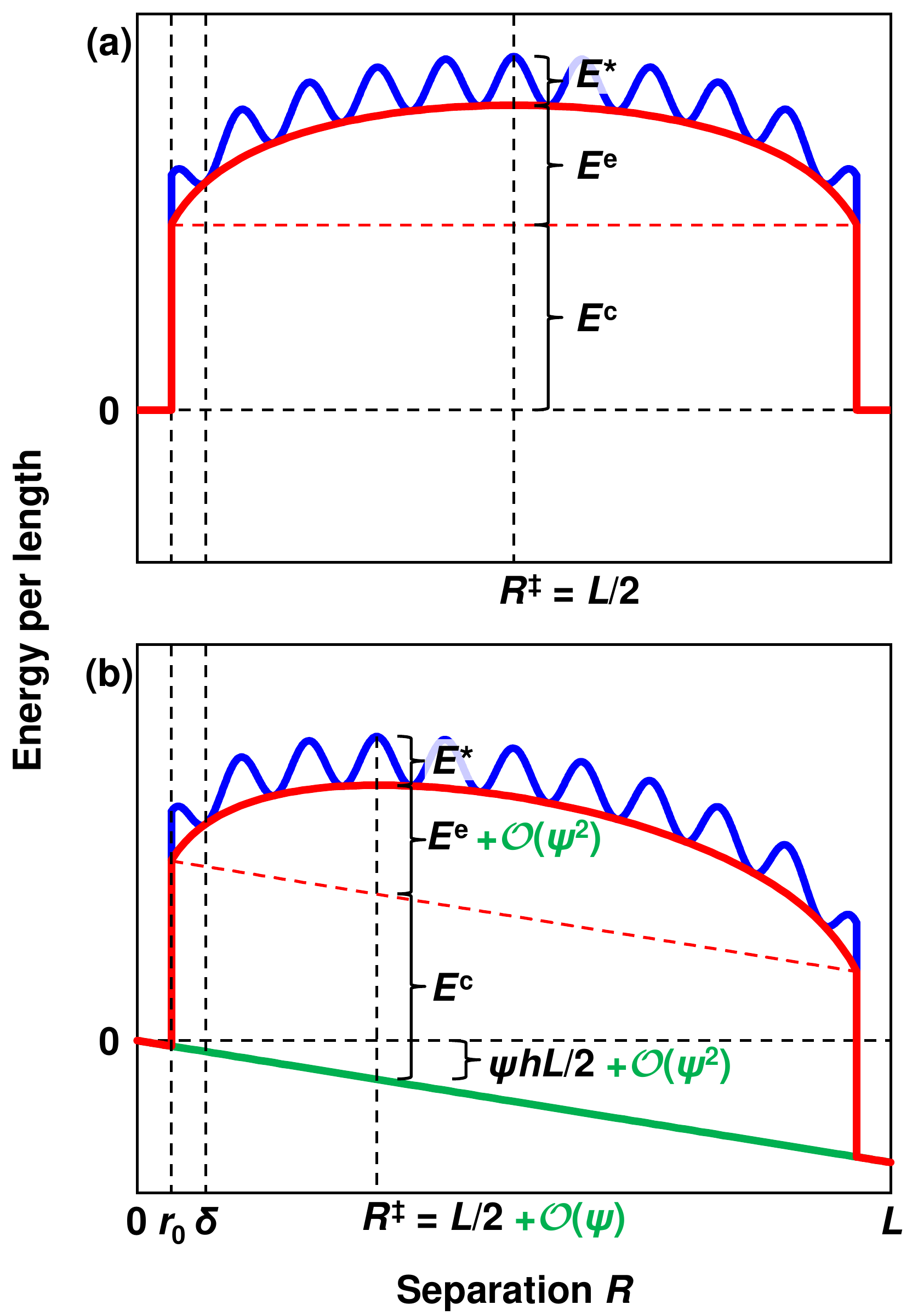}\hspace{-1.78em}%
\caption{\label{energetics}Schematics of the energy vs. separation of a disconnection pair for the cases of (a) $\psi = 0$ and (b) $\psi \ne 0$. 
The red  curves represent $E^\uS(R) = E^\uc + W^\uI(R) + W^\uE(R)$, the blue  curves represent $E(R) = E^\uS(R) + E^\uG(R)$, and the green solid line represents $W^\uE(R) = - \psi h R$. 
$R^\ddagger$ is the ``critical'' separation corresponding to the total energy barrier. 
In (b), the order of the approximation to each part of the energy associated with the external driving force is of order $\mathcal{O}(\psi^2)$. 
}
\end{figure}

The energy change as a function of the separation of the disconnections in a pair by a distance $R$ is illustrated in Fig.~\ref{energetics}. 
The red curve shows that the formation of a disconnection pair (separated by the disconnection core size $r_0$ in one period) requires energy $E^\uc$; this is the disconnection core energy. 
Each disconnection interacts elastically with all other disconnections (including the other disconnection of the same pair and their periodic images). 
To separate a pair of disconnections by distance $R$ against the elastic Peach-Koehler force requires  work $W^\uI(R)$; this is represented by the curved portion of red line in Fig.~\ref{energetics}. 
Expansion of the disconnection dipole against the external driving force $\psi$ also requires work $W^\uE(R)$; this is denoted by the green line in Fig.~\ref{energetics}b. 
In addition, there is a set of energy barriers $E^{\ast}$ resulted from the atomic-scale (Peierls) potential that each disconnection must cross as it migrates. 
Hence, the total change in energy versus disconnection pair separation $R$ is 
\begin{widetext}
\begin{eqnarray}\label{E_R}
E(R) 
&=& E^\uS(R) + E^\uG(R)
= E^\uc + W^\uI(R) + W^\uE(R) + E^\uG(R) 
\nonumber\\
&=& \left\{\begin{array}{ll}
\left(2\gamma |h| + 2\zeta K b^2\right)
+ 2Kb^2 \ln\left|\displaystyle{\frac{\sin(\pi R/L)}{\sin(\pi r_0/L)}}\right| 
- \psi hR + E^\uG(R), & \hspace{3em} r_0 \le R \le L-r_0 \\
0, & \hspace{3em} \text{otherwise}
\end{array}\right.,
\end{eqnarray} 
\end{widetext}
where $E^\uc \equiv 2\gamma|h| + 2\zeta Kb^2$ is an estimate of the disconnection core energy and $E^\uG(R)$ describes the periodic glide barriers which we approximate as $E^\ast [1-\cos(2\pi R/\delta)]/2$. 
Equation~\eqref{E_R} describes the blue curve in Fig.~\ref{energetics}. 

The free energy associated with formation of the disconnection pair should also include the configurational entropy. 
In our model, $L$ is the average spacing between disconnection pairs; hence, $\delta/L$ is the equilibrium concentration of disconnections: 
\begin{equation}\label{equilconc}
\frac{\delta}{L}
= \exp\left(-\frac{E^\uc}{k_\uB T/w}\right).  
\end{equation} 
(See Supplemental Material Section II for more details.)
The configurational entropy (per length) is then
\begin{align}
S 
&= \frac{k_\uB}{\delta w}\left[
L\ln L - \delta \ln \delta - (L-\delta)\ln(L-\delta)
\right]
\nonumber\\
&\approx \frac{k_\uB}{w} \ln\left(\frac{eL}{\delta}\right)
= \frac{k_\uB}{w} + E^\uc/T, 
\end{align}
where the last line applies for $({\delta}/{L})\to 0$
and we applied  Eq.~\eqref{equilconc}. 
Finally, the free energy (per length) as a function of  disconnection separation $R$ may be written as 
\begin{align}
F(R)
&= E(R) - TS
\nonumber\\
&= W^\uI(R) + W^\uE(R) + E^\uG(R) - k_\uB T/w, 
\end{align}
where the individual terms are as per Eq.~\eqref{E_R}.
Recall that the quantities $F$, $E$, $S$, $W^\uI$, $W^\uE$, $E^\ast$, $E^\uc$ and $L$ are dependent on disconnection mode such that in the multi-disconnection mode case, each should have a subscript $m$. 

\subsection*{Kinetics}

We  estimate the GB velocity as the ratio of the step height to the sum of the disconnection (pair) nucleation time $t^\un_m$ and the time required for the disconnections to migrate $t^\ug_m$ the distance required for annihilation $L$: 
\begin{equation}\label{vhtntg}
v=\sum_{m}\frac{h_m}{t^\un_m+t^\ug_m}.
\end{equation}
The most important assumption implied by Eq.~\eqref{vhtntg} lies in the summation over all modes. 
We implicitly make the approximation  that disconnection interactions only occur  between  disconnections of the same mode (of course, this is not true). 
In order to assess this approximation we compare kMC results in Fig.~\ref{fig2}d,   where the red curve shows the GB mobility obtained by adding the velocities of  two, single mode kMC simulations (i.e.,  
$(\tilde{b}, \tilde{h}) = (0,1)$ and $(1,1)$) and the black curve shows the same two modes operating together in one kMC simulation. 
While the agreement is not perfect, it is very good indicating that this is a reasonable approximation. 

We now derive  expressions for $t^\un_m$ and $t^\ug_m$ in Eq.~\eqref{vhtntg}. 
For each mode (temporarily dropping mode index $m$), the free energy barrier for disconnection pair nucleation is $\Delta F = F(R^\ddagger)$, where $R^\ddagger = R^\ddagger(\psi)$ is the driving force-dependent critical separation (i.e., where $F$ is a maximum). 
From Eq.~\eqref{E_R}, 
\begin{equation}\label{DFddagger}
\Delta F
= 2Kb^2 \ln \left[\dfrac{\sin(\pi R^\ddagger/L)}{\sin(\pi r_0/L)}\right] - \psi h R^\ddagger 
+ E^\ast - k_\uB T/w. 
\end{equation}
Expanding $\Delta F$ about $\psi = 0$ (recall that $R^\ddagger$ is a function of $\psi$) and retaining the first order term, we find 
\begin{equation}
\Delta F
= E^\ue+ E^\ast - k_\uB T/w  - \psi hL/2 + \mathcal{O}(\psi^2), 
\end{equation}
where $E^\ue \equiv W^\uI(R=L/2) = - 2Kb^2 \ln[\sin(\pi\delta/2L)]$ represents the contribution of the long-range elastic interactions between disconnections (obviously, for a pure step mode $b = 0$, $E^\ue = 0$). 
The nucleation time is then
\begin{align}\label{tn}
t^\un
&= \frac{1}{r^+ - r^-}
\nonumber\\
&=
\left\{{\omega \exp\left[-\dfrac{\Delta F(h)}{k_\uB T/w}\right] - \omega \exp\left[-\dfrac{\Delta F(-h)}{k_\uB T/w}\right]}\right\}^{-1}
\nonumber\\
&= \frac{1}{2e\omega} 
\exp\left(\frac{E^\ue + E^\ast}{k_\uB T/w}\right)
\csch\left(\frac{\psi hL}{2k_\uB T/w}\right), 
\end{align}
where $r^+$ and $r^-$ represent the rates for GB migration by $h$ and $-h$, respectively, and  $e =\exp(1)$ is Euler's number. 

The energy barrier for disconnection glide over the atomic-scale barriers (i.e., the amplitude of the blue curve in Fig.~\ref{energetics}) is $E^\ast - \psi h\delta/2$. 
The rate of crossing one such barrier is $r^\pm = \omega \exp[-(E^\ast \mp \psi h\delta/2)w/k_\uB T]$. 
Within a period $L$, the number of such glide barriers that must be overcome is $L/2\delta$, such that the time required for disconnections to glide to annihilation is
\begin{equation}\label{tg}
t^\ug 
= \frac{L/2\delta}{r^+ - r^-}= \frac{L}{4\delta\omega} 
\exp\left(\frac{E^\ast w}{k_\uB T}\right)
\csch\left(\frac{\psi h\delta w}{2k_\uB T}\right).
\end{equation}
Again, recall that $t^\un$ and $t^\ug$ in Eqs.~\eqref{tn} and \eqref{tg} depend on disconnection mode such that $t^\un$, $t^\ug$, $E^\ue$, $E^\ast$, $h$, $b$ and $L$ should be assigned a disconnection mode index (subscript) $m$. 

Substituting Eqs.~\eqref{tn} and \eqref{tg} into Eq.~\eqref{vhtntg}, we  obtain an  expression for the GB velocity $v$. 
Then, based on the definition Eq.~\eqref{defM}, the GB mobility is 
\begin{equation}\label{velocity1Model}
M 
= \frac{2\omega \delta w}{k_\uB T}
\sum_m 
\dfrac{h_m^2 \exp\left(-\dfrac{E^\ast_m + E^\uc_m}{k_\uB T/w}\right)}
{1 + \dfrac{2}{e}\exp\left(\dfrac{E^\ue_m - 2E^\uc_m}{k_\uB T/w}\right)}. 
\end{equation}

Equation~\eqref{velocity1Model} is applied to predict the temperature dependence of the GB mobility for each of the   kMC simulation cases in Section~\ref{sec:kMCresults}. 
Compare the theoretical prediction  (blue dashed curves) and the kMC  simulation results (solid black curves) in Fig.~\ref{fig2}. 
Overall, the theoretical predictions from Eq.~\eqref{velocity1Model} capture the major trends in the  kMC simulation data for all cases.
These include 
\begin{itemize}
	\item[(i)] the increase in mobility with increasing temperature at low $T$ (Arrhenius behavior),
	\item[(ii)] the decrease in mobility with increasing temperature at high $T$ (anti-thermal behavior),
	\item[(iii)] the presence of a single mobility peak for the single disconnection mode cases,
	\item[(iv)] the presence of a single or a double peak (and a corresponding minimum) in the mobility for the multi-mode cases, and 
	\item[(v)] the trends in the magnitude of the mobility between different disconnection mode cases.
\end{itemize}

While the statistical disconnection theory reproduces the major trends in the temperature dependence of the mobility, this theory is only semi-quantitative. 
This may be attributed to several approximations in the theory.
These are (1) not including interactions between disconnections of different modes,
(2) assuming the same attempt frequency $\omega$ in the expressions for both of all modes $t^\un_m$ and $t^\ug_m$, 
and (3) the functional form of Eq.~\eqref{vhtntg}.
Writing the denominator in Eq.~\eqref{vhtntg} as the sum of  $t^\un_m$ and $t^\ug_m$  yields the correct solutions in the $t^\un_m \gg t^\ug_m$ and $t^\un_m \ll t^\ug_m$ limits, but is only approximate  between these limits.

\subsection*{Trends in $M(T)$}

We examine several of the main features of the temperature dependence of the mobility based on the statistical disconnection model.

In the high temperature limit, each term in the summation of Eq.~\eqref{velocity1Model} will have the form $h^2_m \exp(-|h_m|)$, which  converges rapidly to zero with increasing $h_m$ (or $m$). 
This suggests that at high temperature the summation can be truncated; yielding a constant. 
This then implies that $M \sim 1/T$ in the  high temperature limit. 

\begin{figure}[hb]
\includegraphics[height=0.8\linewidth]{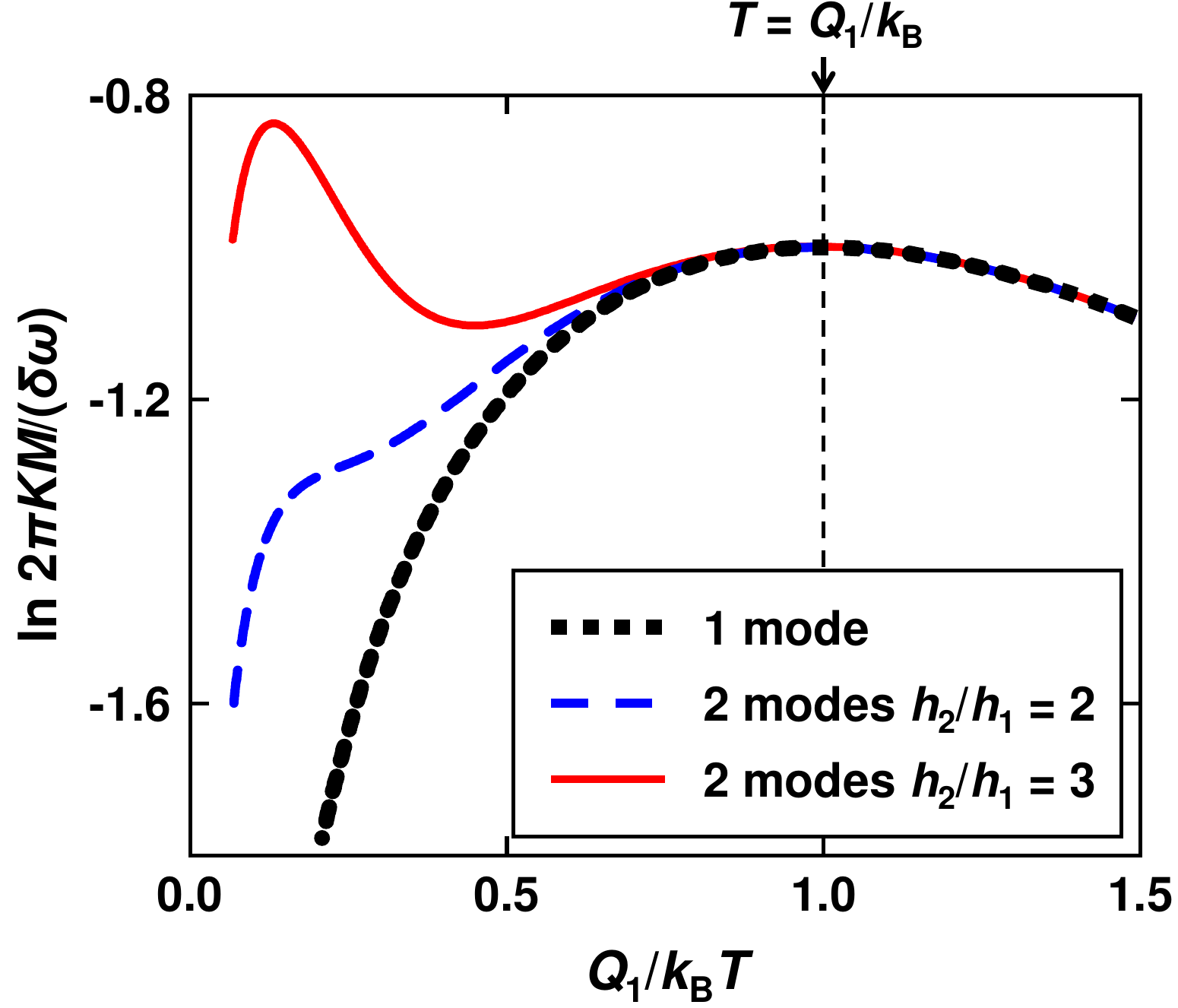}\hspace{-1.78em}%
\caption{\label{AnalysisModel}Temperature dependence of GB mobility described by Eq.~\eqref{velocity2Model} for a single mode (black dotted curve), two modes with $h_2/h_1 = 2 < e$ (blue dashed curve), and two modes with $h_2/h_1 = 3 > e$ (red solid curve) cases. In the two-mode simulations $Q_2=10Q_1$.}
\end{figure}

In the low temperature limit ($T \to 0$), each term in the summation in Eq.~\eqref{velocity1Model} is $(eh_m^2/2)\exp[-(E^\ast_m + E^\ue_m - E^\uc_m)/(k_\uB T/w)]$ for  $E^\ue_m > 2E^\uc_m$ and  $h_m^2 \exp[-(E^\ast_m + E^\uc_m)/(k_\uB T/w)]$ for $E^\ue_m < 2E^\uc_m$. 
In other words, each term in the summation in Eq.~\eqref{velocity1Model} is of the form $(A_m h_m^2/k_\uB T) \exp(-Q_m/k_\uB T)$, where $A_m$ and $Q_m$ are positive constants (the detailed forms of which depend on the relative magnitudes of $E^\ue_m$, $E^\uc_m$ and $E^*_m$). 
This implies that the GB mobility has the form 
\begin{align}\label{MlowT}
\ln M 
&= \ln\sum_m \frac{A_m h_m^2}{k_\uB T} \exp\left(- \frac{Q_m}{k_\uB T}\right)\nonumber\\
&\approx \ln\left[\frac{A_1 h_1^2}{k_\uB T} \exp\left(- \frac{Q_1}{k_\uB T}\right) \right]\nonumber\\
&\approx -\frac{Q_1}{k_\uB T}\left[1 + \mathcal{O}(T\ln T)\right]
\quad \text{as $T \to 0$}. 
\end{align}
If the $m=1$ disconnection mode corresponds to the minimum $Q_m$ amongst all the modes, we need only consider this mode in the summation (Eq.~\eqref{MlowT}, second line). 
Equation~\eqref{MlowT} implies that, as $T \to 0$, the GB mobility is Arrhenius  (Eq.~\eqref{MlowT}, last line). 

When $E^\ue_m \ll E^\uc_m$ (e.g., for a pure step),
\begin{equation}\label{velocity2Model}
M = \frac{2\omega\delta w}{k_\uB T}
\sum_m h_m^2 \exp\left(-\frac{Q_m}{k_\uB T}\right), 
\end{equation}
over a wide temperature range ($Q_m = (E^\ast_m + E^\uc_m)w$). 
Figure~\ref{AnalysisModel} shows the temperature dependence of the GB mobility in this limit in classical Arrhenius coordinates.
For a  single mode (black dotted curve), the GB mobility increases and then decreases with increasing  temperature (the mobility exhibits a maximum with respect to temperature).
At low temperature, the mobility is Arrhenius,  but decreases with increasing $T$ at high temperature - this is the so-called  anti-thermal behavior \cite{Homer2014}.
The temperature at which this behavior switches is roughly given by $\ud M/\ud T = 0$ or  $T^\uc = Q_1/k_\uB$ using Eq.~\eqref{velocity2Model}. 
Similar conclusions apply when $E^\ue_m \gg E^\uc_m$ 

We now examine why there is a local minimum in  the $M(T)$ kMC data for some multi-disconnection mode simulations (Fig.~\ref{fig2}e,f).
Consider two disconnection modes, $m = 1$ and $2$ in  Eq.~\eqref{velocity2Model} as indicated by the solid red ($h_2/h_1=3$) and blue dashed ($h_2/h_1=1$) curves in Fig.~\ref{AnalysisModel}.
When $Q_2\gg Q_1$, the critical temperature (corresponding to  $m = 2$) occurs at a much higher temperature than for the  the $m = 1$ mode.
This implies that there may be a local minimum in the $M$ vs. $T$ data, provided that the two peaks are of sufficient height. 
That is, a local minimum exists for $Q_2\gg Q_1$ if and only if $h_2/h_1>e$ (see Eq.~\eqref{velocity2Model}). 
This is in qualitative agreement with the kMC results in Fig.~\ref{fig2}, where a local minimum exists when $h_2/h_1> 2$ and gets deeper as $h_2/h_1$ increases.
As a specific example, consider the  $\Sigma 25$ $[100]$ $(034)$ STGB examined in the MD simulations of Homer, et al.~\cite{Homer2014}, where $h_2/h_1=25/7>e$ such that $M(T)$ should exhibit a local minimum (provided it occurs  below the melting point); indeed, this is consistent with the MD data where a local minimum is observed for this GB in Ni, but not in Al.
We suspect that the GB mobility in Al too would show a local minimum, if it did not melt first (note, the  melting point of Ni is nearly twice that of Al).

\section{Parameters for specific grain boundaries and comparison with MD}\label{MDcomparison}

While the kMC simulations described in Section~\ref{sec:kmc}  employed parameters chosen to investigate the general features of the temperature dependence of GB mobility, the kMC  can also be applied to simulate the motion of a specific GB in a real material - provided  the appropriate GB parameters are available. 
Here, we develop a parameters set for the $\Sigma 17$ $[100]$ $(035)$ and $\Sigma 25$ $[100]$ $(034)$ STGBs in EAM Al~\cite{Mishin1999}, perform kMC simulations for these two specific GBs, and compare the GB mobility thus obtained with molecular dynamics results.

\begin{table*}[t]
\newcolumntype{Y}{>{\centering\arraybackslash}X}
\renewcommand\arraystretch{1.5}
\caption{\label{tab:parameters}
Parameters for two symmetric tilt GBs in EAM\cite{Mishin1999} Al. 
$a_0 = 0.405$~nm is the lattice constant, 
$a_\text{dsc}$ is the size of a DSC lattice cell, 
$m$ is the index of disconnection mode, 
$b_m$, $h_m$ and $E_m^\ast$ are the the corresponding Burgers vector, step height, and glide barrier,
$A_m$ and $B_m$ are  fitting parameters (see Eq.~\eqref{AmBm}),
$K \equiv \mu/[4\pi(1-\nu)]$ ($\mu$ and $\nu$ are  the shear modulus and  Poisson's ratio),
$\gamma$ is the step energy ,
$\delta_0$ is the effective core size ($r_0$ is the core size),
$\delta$ is the spacing between lattice point in the quasi-1D lattice model, and
$\zeta$  scales the contribution of the core energy (Eq.~\eqref{E_R}). 
}
\begin{ruledtabular}
\begin{tabular}{cccccccccccc}
$m$ & $b_m/a_\text{dsc}$ & $h_m/a_\text{dsc}$ & $A_m$ (eV/nm) & $B_m$ (eV/nm) & $E_m^\ast$ (eV/nm) & $K$ (eV/nm$^3$) & $\gamma$ (eV/nm$^2$) & $\pi\delta_0/L$ & $\pi\delta/L$ & $\zeta$ & $r_0/\delta$ \\
\hline
\multicolumn{10}{l}{$\Sigma 17$ $[100]$ $(035)$ with $a_\text{dsc} = a_0/\sqrt{34}$}\\
1 & 0 & $8.5$ & $1.07570$ & $0$ &
$1.24444$ &
\multirow{4}{*}{$20.1013$} & 
\multirow{4}{*}{$1.0656$} & 
\multirow{4}{*}{$0.28513$} &
\multirow{4}{*}{$0.31416$} &
\multirow{4}{*}{$-1.903$} &
\multirow{4}{*}{$0.134$}\\
2 & 1 & $-2$ & $0.47855$ & $0.18376$ & $0.19753$ &  & \\
3 & 1 & $6.5$ & $1.39769$ & $0.12727$ & $1.11111$ & & \\
4 & 2 & $-4$ & $1.56834$ & $0.79501$ & $0.29876$ & & \\
\multicolumn{2}{l}{Estimates} & & & & & $20.34$-$24.53$\footnote{\label{footnotea}The range of $K = \mu/[4\pi(1-\nu)]$ correspond to different  crystal orientations and the anisotropic elastic constants are  from Ref.~\onlinecite{Mishin1999}. } 
& $1.3844$\footnote{The $\Sigma 17$ $[100]$ step energy is estimated from $\sqrt{2}\gamma_\us - \gamma_0$, where $\gamma_\us = 0.4861$~J/m$^2$ for the $(014)$ and $\gamma_0 = 0.4657$ J/m$^2$ for the $(035)$ STGB in EAM Al; $\gamma_0$ and $\gamma_\us$ are  from Ref.~\onlinecite{Han2017}. } \\
\hline
\multicolumn{10}{l}{$\Sigma 25$ $[100]$ $(034)$ with $a_\text{dsc} = a_0/\sqrt{25}$}\\
1 & 0 & $12.5$ & $3.25669$ & $0$ &
$0.99012$ &
\multirow{5}{*}{$19.2861$} & 
\multirow{5}{*}{$1.6282$} & 
\multirow{5}{*}{$0.06038$} &
\multirow{5}{*}{$0.15708$} &
\multirow{5}{*}{$-1.044$} &
\multirow{5}{*}{$0.135$}\\
2 & 1 & $-3.5$ & $0.79239$ & $0.24639$ & $0.42222$ &  & \\
3 & 1 & $9$ & $3.21678$ & $0.2472$ & $1.14568$ & & \\
4 & 2 & $5.5$ & $4.54659$ & $0.95765$ & $1.17778$ & & \\
5 & 3 & $2$ & $6.89744$ & $2.30333$ & $1.20741$ & & \\
\multicolumn{2}{l}{Estimates} & & & & & $20.34$-$24.53^{\ref{footnotea}}$
& $1.7383$\footnote{The $\Sigma 25$ $[100]$ step energy estimates employ $\gamma_\us = 0.4707$~J/m$^2$ for the $(017)$ and $\gamma_0 = 0.3872$ J/m$^2$ for the $(034)$ STGB (see footnote b).} \\

\end{tabular}
\end{ruledtabular}
\end{table*}

\begin{figure}
\includegraphics[height=1.08\linewidth]{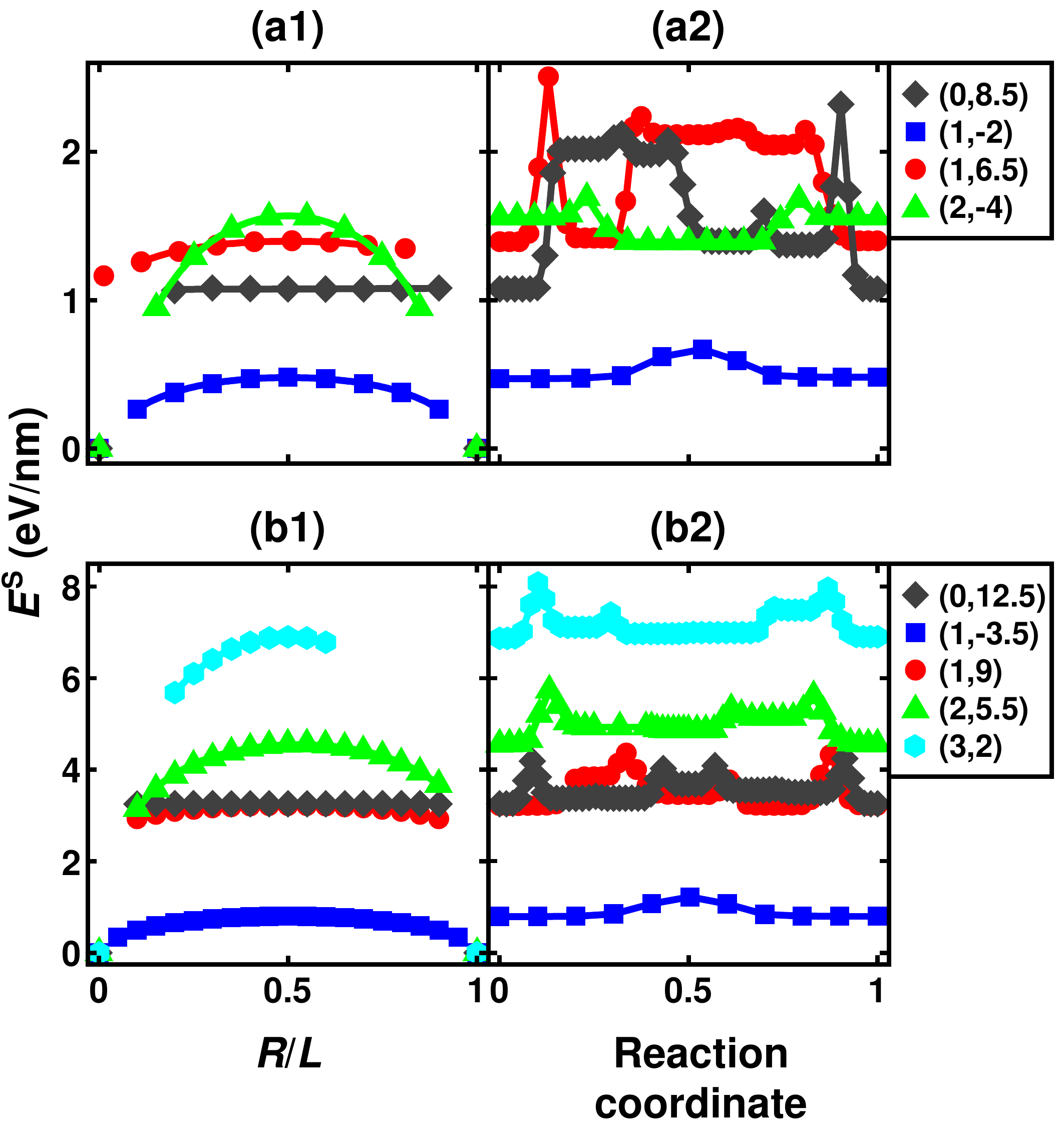}
\caption{\label{NEB}(a1) and (b1) are the energy $E^\uS$ vs. separation $R$  of a pair of disconnections, scaled by the supercell period $L$ for the $\Sigma 17$ $[100]$ $(035)$ and $\Sigma 25$ $[100]$ $(034)$ STGBs in EAM Al, respectively. 
The data point colors indicate disconnection modes $(b_m/a_\text{dsc}, h_m/a_\text{dsc})$, as indicated in the legend.
(a2) and (b2) are the energy landscape between two neighboring metastable configurations with the highest energy (at the maxima of each curve in (a1) and (b1)) obtained from the NEB calculations. 
}
\end{figure}

The parameters required for the kMC simulations are $K$, $\gamma$, $\zeta$, $\delta$, $\omega$, $\{b_m, h_m\}$ and $\{E_m^\ast\}$. 
For any coincidence-site-lattice GB of  particular bicrystallography, $\{b_m, h_m\}$ can be determined based on the GB geometry\cite{Han2018}. 
The parameters $K$, $\gamma$, $\zeta$ and $\delta$ can be determined  from $E_m^\uS$ vs. $R$ with $\psi = 0$ (see the red curve  in Fig.~\ref{energetics}), which we determine via molecular static  for each disconnection mode for each GB.  
For each mode, we constructed a series of configurations where a pair of disconnections were separated by different distance $R$. 
Then, we minimized the energy of each of the prepared configurations and obtained a set of metastable states. 
The results are plotted in Figs.~\ref{NEB}a1 ($\Sigma 17$) and \ref{NEB}b1 ($\Sigma 25$); each data point corresponds to a metastable state after energy minimization. 
Based on the molecular statics data of each mode, we extract the parameters, as described below. 

\begin{itemize}
\item[(i)] 
Fit $E^\uS_m$ vs. $R$ to 
\begin{equation}\label{AmBm}
E^\uS_m(R) = A_m + B_m\ln\sin[\pi(R/L + c_m)], 
\end{equation}
where $A_m$, $B_m$ and $c_m$ are the fitting parameters. 
The fitted curves are plotted in Figs.~\ref{NEB}a1,b1. 
Comparing Eqs.~\eqref{AmBm} and \eqref{E_R}, we see $B_m = 2Kb_m^2$. 
$K$ is obtained by fitting $\{B_m\}$ and $\{b_m^2\}$ (for all modes). 

\item[(ii)] 
Comparing Eqs.~\eqref{AmBm} and \eqref{E_R}, we see that
\begin{equation}\label{Am}
A_m = 2\gamma |h_m| - B_m \ln\left[e^{-\zeta} \sin(\pi r_0/L)\right]. 
\end{equation}
(We assume that $r_0$ is the same for all modes.) 
We denote 
$C_{m1} \equiv 2|h_m|$, 
$C_{m2} \equiv -B_m$, 
$X_1 \equiv \gamma$ 
and $X_2 \equiv \ln\left[e^{-\zeta} \sin(\pi r_0/L)\right]$. 
Then, Eq.~\eqref{Am} becomes
\begin{equation}\label{X1X2Y}
C_{m1} X_1 + C_{m2} X_2 = A_m 
\quad\text{or}\quad 
\mathbf{C}\mathbf{X} =\mathbf{A}. 
\end{equation}
We obtain $X_1$ and $X_2$ by fitting Eq.~\eqref{X1X2Y} to the  $\{C_{m1}, C_{m2}, A_m\} \equiv \{2|h_m|, -B_m, A_m\}$ data. 
Minimization of $|\mathbf{CX}-\mathbf{A}|$ gives $\mathbf{X} = (\mathbf{C}^T\mathbf{C})^{-1}(\mathbf{C}^T\mathbf{A})$. 
From this, we find  the step energy $\gamma = X_1$. 

\item[(iii)] 
Defining $\sin(\pi\delta_0/L) \equiv e^{-\zeta}\sin(\pi r_0/L) = e^{X_2}$ ($\delta_0$ is the effective disconnection core size), 
we  obtain $\delta_0$ from $\pi\delta_0/L = \arcsin e^{X_2}$. 
However, the kMC simulations require  $\zeta$ and $\delta$. 
As discussed above, these parameters are related by $\zeta + 2 = \ln(\zeta/\zeta_0)$; as long as this relationship is satisfied  the lattice model and continuum theory will be consistent (see Supplemental Material Section III). 
In practice, we choose $\delta$ as the size of a CSL cell, such that $\zeta = \ln(\delta/\delta_0) - 2$. 
\end{itemize}

We obtain the parameter $E_m^\ast$ for each mode from atomistic data obtained using the nudged-elastic-band (NEB) method~\cite{Jonsson1998,Henkelman2000}. 
To do this, we first choose two neighboring states corresponding the the minimum energy configurations near the top of the profile (see the blue curve in Fig.~\ref{energetics}a); these correspond to the initial and final states in the NEB calculations. 
The NEB method finds the  minimum energy path (reaction coordinates) associated with the transition from the initial (reaction coordinate 0) to the final (reaction coordinate 1) states.
These data are shown in Figs.~\ref{NEB}a2  ($\Sigma 17$) and \ref{NEB}b2 ($\Sigma 25$) for several of the lowest energy disconnection modes. 

The parameters obtained by fitting the molecular statics and NEB results for the two GBs are summarized in Table~\ref{tab:parameters}. 
The elastic modulus $K$ obtained from fitting is close to the expected value for a perfect crystal. 
The excess energy associated with the introduction of a step can be estimated as $\gamma = \sqrt{2}\gamma_\us - \gamma_0$, where $\gamma_0$ is the energy of the flat GB and $\gamma_\us$ is the energy of a GB with inclination of $45^\circ$. 
The value of $\gamma$  obtained by fitting is consistent with this estimated value. 
Hence, the fitting based on Eq.~\eqref{AmBm} is reasonable. 
While $\zeta<0$ in Table~\ref{tab:parameters} seems counter-intuitive, we recall that $\zeta$ is the scaling factor for the core energy only if the chosen core size $\delta/2$ is the real disconnection core size.
Since we fix $\delta$ as the size of a CSL cell (probably too large), this core energy has no physical meaning. 
Rather, we choose the core energy parameter and core size self-consistently, as in classical dislocation theory. 

\begin{figure}[ht]
\includegraphics[height=1.2\linewidth]{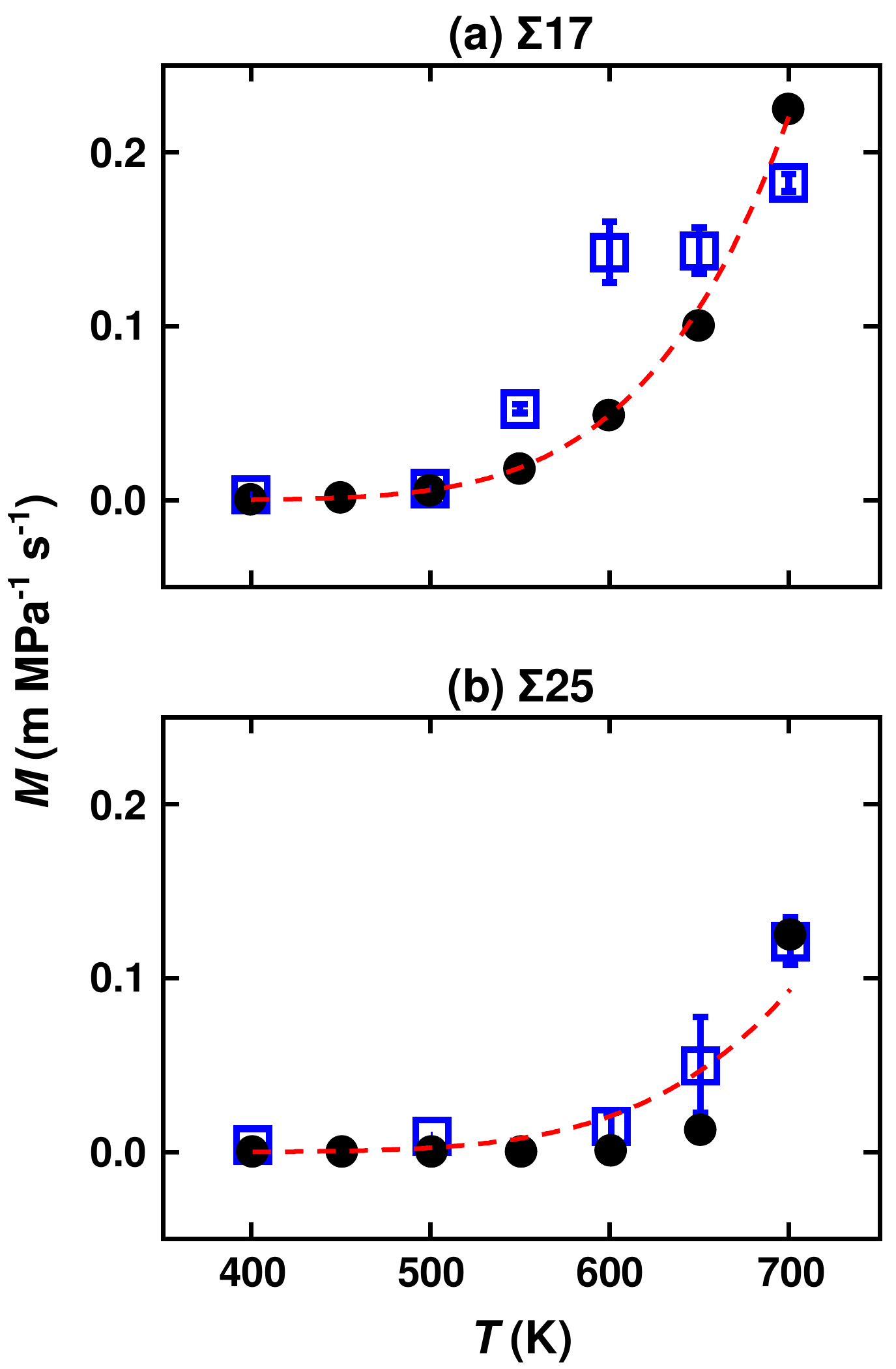}\hspace{-1.78em}%
\caption{\label{MDcompare}Temperature dependence of GB mobility obtained by MD simulations (blue hollow squares), kMC simulations (black solid circles) and disconnection theory (red dashed lines) for (a) $\Sigma 17$ $[100]$ $(035)$ and (b) $\Sigma 25$ $[100]$ $(034)$ STGBs in EAM Al. The error bars in the MD data represent the mobilities obtained in two MD simulations and the hollow, blue squares represent their mean. 
}
\end{figure}

With these atomistically-determined parameters (summarized in Table~\ref{tab:parameters}), we performed kMC simulations for the $\Sigma 17$ $[100]$ $(035)$ and $\Sigma 25$ $[100]$ $(034)$ STGBs in EAM Al and determined  the GB mobility vs. temperature (see the black solid circles in Fig.~\ref{MDcompare}). 
To test our kMC approach, we also determine the GB mobility as a function of temperature from  MD simulations of GB migration (synthetic driving force) using the Large-scale Atomic/Molecular Massively Parallel Simulator (LAMMPS) \cite{Plimpton1995} and the same interatomic potentials (as  used to determine the kMC parameters).
Note that since the kMC model (and theory) are quasi-1D, we performed the bicrystal MD simulations using a simulation cell that is very small in the direction parallel to the tilt axis ($w \sim 1.2$~nm). 
The other two cell dimensions are $\sim 20$~nm (parallel to the tilt axis) and $\sim 100$~nm (orthogonal to the GB plane).
All simulations (see [\onlinecite{Chen2019}] for more details) were run for $3.5$~ns and $\psi=10$~MPa at fixed number of atoms and temperature. 
The GB position is defined as the $x_3$ position where the layer-averaged centro-symmetry parameter~\cite{Stukowski2010} is maximum~\cite{Kelchner1998}.
The GB migration velocity is the normal velocity of the mean GB plane.
The MD results are shown  as the  hollow  blue squares in Fig.~\ref{MDcompare}.  


Figure~\ref{MDcompare} shows that our MD and kMC simulation results are in very good agreement with each other (note, the attempt frequency $\omega$ in the kMC is not determined by our atomistic simulations - here, we chose it on a best fit basis). 
Figure~\ref{MDcompare}  also shows (red dashed curve) $M(T)$ based upon the disconnection theory (Eq.~\eqref{velocity1Model} with the same parameters as for the kMC).
The statistical disconnection model reproduces both the kMC and MD results. 
We note that while the MD simulations involve the fewest assumptions in the prediction of the GB mobility and its temperature dependence, it is also the most computationally costly and impractical at low temperature when the GB mobility is not fast. 
On the other hand, accurate kMC and theoretical predictions of $M(T)$ require determination of the fundamental parameters (such as the energy landscape for the motion of disconnections) - these too can require substantial computational resources (albeit much less than the MD simulations for systematic investigations).
The greatest advantage of both the kMC and statistical disconnection theory lies in their utility for systematic investigation of GB migration physics.

\section{Discussion and Conclusions}

The aim of this paper is to provide a mechanistic understanding of the diverse forms of the  temperature dependence of the grain-boundary (GB) mobility that were previously reported based upon MD  simulations and experiments.  
Several forms of this temperature dependence have been particularly perplexing, in light of the established theories of how GBs migrate (particularly,  ``anti-thermal'' behavior where the GB mobility decreases with increasing temperature).

This paper provides two approaches for understanding $M(T)$ based upon a disconnection-based mechanism for GB migration: a kinetic Monte Carlo implementation and a statistical disconnection theory. 
The main idea employed here is that GB migration occurs via the formation and glide of disconnections along the GB; while disconnection dynamics depends on both the disconnection dislocation vector and step character $(b,h)$, GB migration itself takes place through the motion of the step-component of the disconnection motion. 
We investigate how different disconnection modes give rise to different GB mobilities and how the simultaneous migration of different disconnection modes explains some of the richness of the temperature dependence of the GB mobility.

One of the main results of the present work is that the temperature dependence of the GB mobility is related to the fundamentally different disconnection dynamics at low and high temperature.
In general, GB migration controlled by a single disconnection mode will lead to an Arrhenius $T$-dependent mobility at low temperature and a mobility at high $T$ that is inversely proportional to temperature.
Depending on the disconnection migration energy landscape, the high temperature regime may not be observed (i.e., if the transition temperature is higher than the melting temperature.
The transition between the low and high temperature regimes occurs at different temperatures in different GBs within one material and in different materials for the same GB bicrystallography. 
The transition or critical temperature can be estimated as  $Q_1/k_\uB$, where $Q_1$ is the activation energy for the most easily activated disconnection mode. 
It is the high temperature behavior, where the GB mobility scales inversely with temperature that is responsible for the reported ``anti-thermal'' behavior. 
Both the kMC and theory capture this behavior.

GB migration may also be affected by the activation of multiple disconnection modes; this depends on the relative formation and migration energy of the lowest energy disconnection modes.
Activation of two disconnections, can give rise to GB mobilities versus temperature that exhibit two maxima and a local minimum (as seen in the kMC, theory and experiments). 

While our main focus has been to understand the mechanistic origin of the factors that affect the temperature of the GB mobility, we also predict $M(T)$ for two specific grain boundaries in aluminum. 
To do this, we performed a series of static relaxation calculations for these two GBs as well as nudged elastic band calculations to determine the energy landscape associated with disconnection motion in a material described using an EAM interatomic potential. 
These calculations provided the parameterization for both the kMC and statistical disconnection dynamics theory.
We then compared the kMC and theory with a series of molecular dynamics simulations of the migration of these GBs.
The MD, kMC and theory are all in good agreement. 
This implies that both the kMC and theory can be used to qualitatively predict GB migration behavior. 
While such parameterization of the kMC and theory via atomistic calculations is time consuming, the computational burden is much lower than the full MD simulations when the goal is systematic prediction (e.g., $T$-dependence, bicrystallography-dependence, ...).

Although we have discussed the existence of different temperature regimes for the GB mobility and different forms of $M(T)$ as more disconnection modes are activated, we note that there are additional physical phenomena that may have a profound  effect on the GB mobility.  
Perhaps, the most important of these, not included here, are GB structural phase transitions (also known as complexion transitions~\cite{Cantwell2014}).
While GB structural phase transitions relate to the underlying structure of (even) flat GBs without disconnections, the varied temperature dependences of the GB mobility discussed here is the result of GB dynamics related to disconnection motion.  
Disconnection dynamics can also give rise to finite temperature phase transitions as well, but such transitions do not (necessarily) change the  structure of the underlying GB itself.

\begin{acknowledgements}
This research was  sponsored by the Army Research Office and was accomplished under Grant Number W911NF-19-1-0263. The views and conclusions contained in this document are those of the authors and should not be interpreted as representing the official policies, either expressed or implied, of the Army Research Office or the U.S. Government. The U.S. Government is authorized to reproduce and distribute reprints for Government purposes notwithstanding any copyright notation herein.
\end{acknowledgements}

\bibliography{mybib}

\end{document}